\documentclass{article}

\usepackage[preprint]{neurips_2023}

\usepackage[utf8]{inputenc} 
\usepackage[T1]{fontenc}    
\usepackage{hyperref}       
\usepackage{url}            
\usepackage{booktabs}       
\usepackage{amsfonts}       
\usepackage{nicefrac}       
\usepackage{microtype}      
\usepackage{xcolor}         
\usepackage{graphicx}

\title{Explore the possibility of advancing climate negotiations on the basis of regional trade organizations: A study based on RICE-N}

\author{%
  Wubo Dai\\
  School of Smart Energy\\
  Shanghai Jiao Tong University\\
  Shanghai, 200240, China \\
  \texttt{Dai.WB@sjtu.edu.cn} \\
}

\begin{document}

\maketitle

\begin{abstract}
Climate issues have become more and more important now. Although global governments have made some progress, we are still facing the truth that the prospect of international cooperation is not clear at present. Due to the limitations of the Integrated assessment models (IAMs) model, it is difficult to simulate the dynamic negotiation process. Therefore, using deep learning to build a new agents based model (ABM) might can provide new theoretical support for climate negotiations.
Building on the RICE-N  model, this work proposed an approach to climate negotiations based on existing trade groups. Simulation results show that the scheme has a good prospect.
\end{abstract}

\section{Introduction}

Climate change has significant impacts on the natural environment and human society. Besides, the greenhouse gases causing climate change do not respect national boundaries, their effects can be felt worldwide. Therefore, collective action is required to address this global issue effectively. Although we have made some progress such as the Paris Agreement and the recent COP 27 meets, according to the latest IPCC 6 report [1], we are not doing enough. The report [1] also notes that international cooperation is a critical enabler for accelerated climate action. However, the prospect of international cooperation is not clear at present. According to Nordhaus‘s research [2], the main reason for the hindrance lies in the behavior of "free-ride", that is, countries have an incentive to rely on the emissions reductions of others without taking proportionate domestic abatement.

In order to use AI to provide a solution to this problem, the organizers used Nordhaus‘s RICE model [3] as the climate and economic dynamics framework, combined with ABM, created this RICE-N model that included 27 agents and their negotiation mechanism [4]. Building on the official model, we adopted a form similar to the customs unions, with added the requirement for mitigating rates. Simulation results show that the scheme has a good prospect.

\section{Methodology}
\label{Methodology}

Previous researchers have proposed a lot of solutions to the climate negotiations. The Nordhaus Climate Club [2] is considered to be a very promising climate negotiation mechanism, so our research is based on this theory and real-world situations. Figure 1 illustrates the framework of the default RICE-N model [4] and the model of this work.

\begin{figure}
  \centering
  \includegraphics [width=0.7\textwidth]{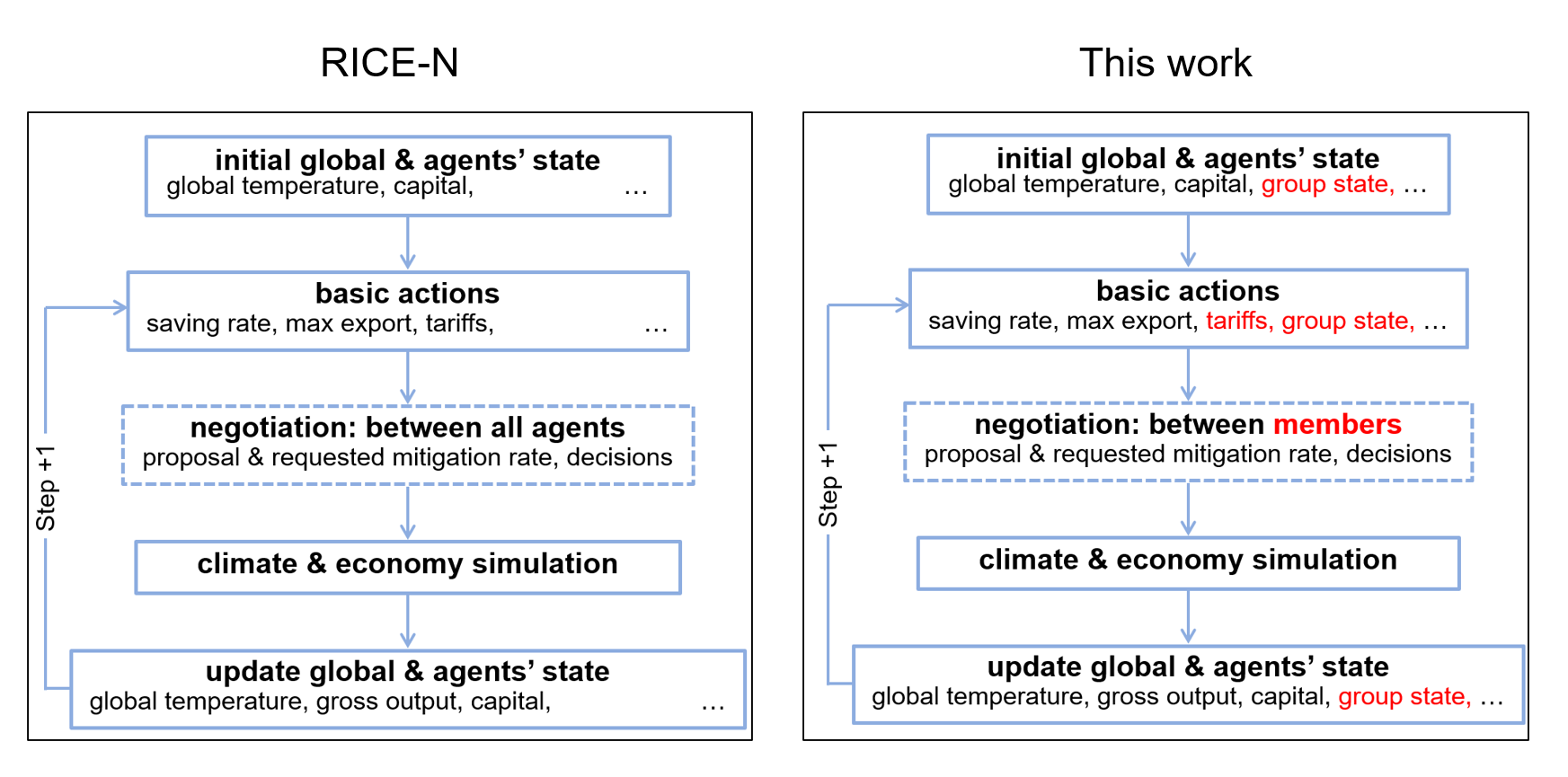}
  \caption{Model structure of RICE-N and this work.}
\end{figure}

\paragraph{Clubs} The first thing to do is to build clubs, we set a new parameter (an int number) to represent the state of the group the agents are in at each step, and allow them to choose their group state freely when a new step begin, including setting up / joining in / changing or quitting a group. For example, assume that the grouping state of agents A and B in time step one are 1 and 2, that means they were in the different group now. If agent A still stay in group 1 in time step two, then agent B can choose its grouping state to 1, that means they will stay in the same group in this time step. Besides, B could choose its grouping state to 0 (do not attend any group), 2 (stay in the previous group) or 3(create a new group, assume no agent's state is 3 in last step).

In Nordhaus’s theory [2], the foundation of the club is a comparable carbon-pricing mechanisms. Considering that countries cannot reach a consensus on the carbon pricing mechanism at present, we choose to use the default tariff-mitigation rate negotiation mechanism here. However, the negotiation will only be carried out among members of the same group.

\paragraph{Motivation} The motivation of the club has been described as sanction to non-members and tariff-free border between members [2]. One of the sanction ways is a countervailing duty on the carbon content of imports, just like the Carbon Border Adjustment Mechanism of the EU [5]; However, it might be difficult to achieve in the model now, so we chose another way Nordhaus mentioned: an extra tariff on all imports from non-members. 

As for tariff-free border between members, after referring to the situation of some trade organizations, such as: North American Free Trade Area (NAFTA), China-ASEAN Free Trade Area (CAFTA) [6], we found that their zero tariff settings are gradually realized, so here we draw on this way, tariff for members will decrease to zero over time.

\paragraph{Critical mass} Another important principle in climate clubs is called critical mass, which means it needs the participation of the major economic players, as they represent 61\% of global gross domestic product and 43\% of goods imports [7]. This means we need to set an in initial group state to get these countries in one club at the beginning. However, the real situation is it is difficult for them to reach an agreement in one step, considering the complex international relations.For example, China and the US ‘s ambiguity to the EU’s Carbon Border Adjustment Mechanism [8].

Based on these situations, we propose a concept: the major economic players do need participate but will separately at the beginning; and the beginning group state will be based on existing trade cooperation to make the negotiation easier. And hopefully, this will finally lead to climate club of global scope.

To simulate this scenario, we ranked 27 agents according to their carbon emissions and productions, The initial data comes from the previous work of Tianyu et al [4]. Then we divided agents into high carbon emitting countries (HC) and low carbon emitting countries (LC). You may see the {\bf Supplementary Mater 1} for the detail calculation. Finally, we have two different initial group state:

For {\bf HCs}: we put the top five agents into one group, randomize other agents into four other groups to simulate the ideal state where HC reach an agreement at the beginning.

For {\bf HC with LC}: we put the top five agents into five different groups, put other agents into them evenly based on the rank to simulate the state mentioned above.

\paragraph{Real world situation} There are some other sets to close to real world situation, like for saving rate and max export, choices are limited based on World Bank global average data. You may see the {\bf Supplementary Mater 2} for the detail explanation.

\section{Results and Discussion}
\label{Results and Discussion}

The temperature rise and total gross output during 100 years in the simulation under two scenarios are shown in Figure 2. From the figure, we can find that the temperature rise of the HC with LC group is lower than that of the HCs by more than 0.8 °C when the gross output has almost no difference. 

\begin{figure}
  \centering
  \includegraphics [width=0.7\textwidth]{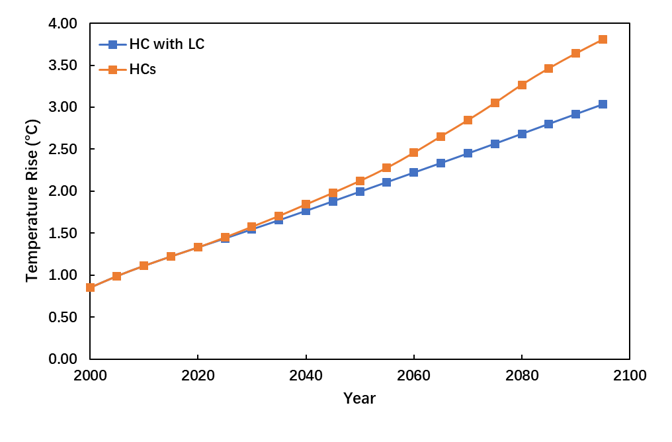}
  \includegraphics [width=0.7\textwidth]{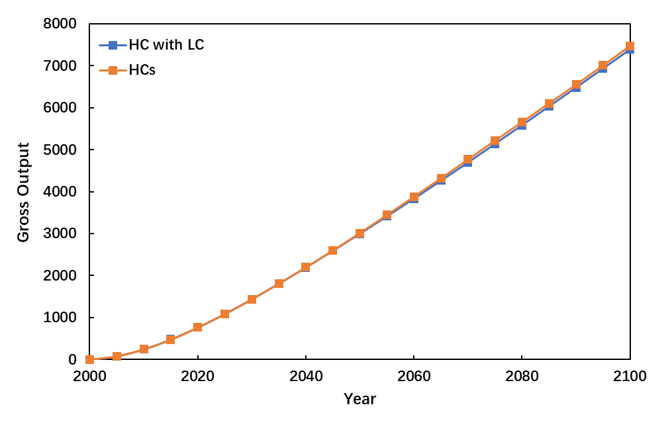}
  \caption{temperature rise and total gross output.}
\end{figure}

To find out how this discrepancy aroused, We averaged the mitigating rates of all agents in each stage, and displayed the data in Figure 3. According to the Average mitigating rate of all regions, We found that in HCs, average mitigating rate was lower than HC with LC in some periods, which led to a higher temperature rise. 
 
\begin{figure}
  \centering
  \includegraphics [width=0.7\textwidth]{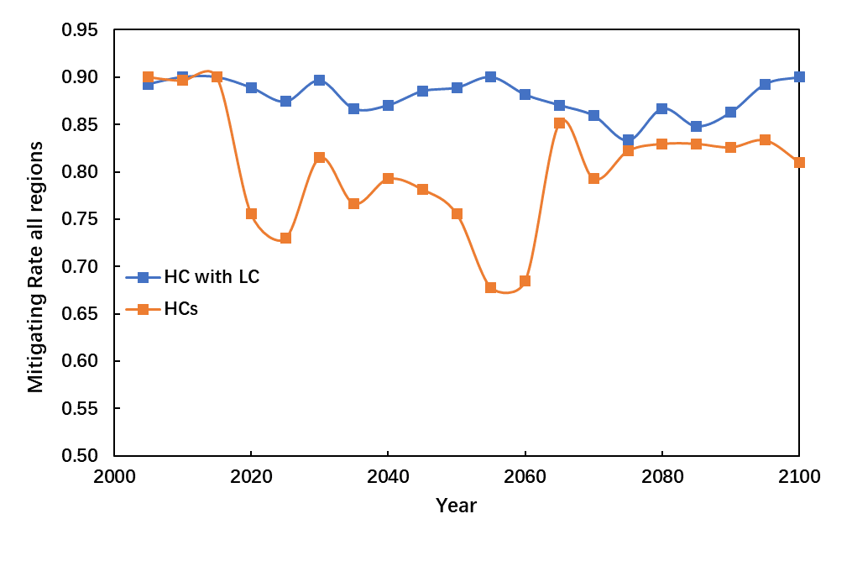}
  \caption{average mitigating rates of all agents.}
\end{figure}
 
Another thing we found was that according to the final group state, in HC with LC, agents tended to cluster together at the end than the other scenario. It also included more high carbon emitting agents.That was closer to the ideal eventuality of a globally unified climate club. The results are shown in Figure 4. However, There are still many aspects to explore and improve about this research.

\begin{figure}
  \centering
  \includegraphics [width=0.7\textwidth]{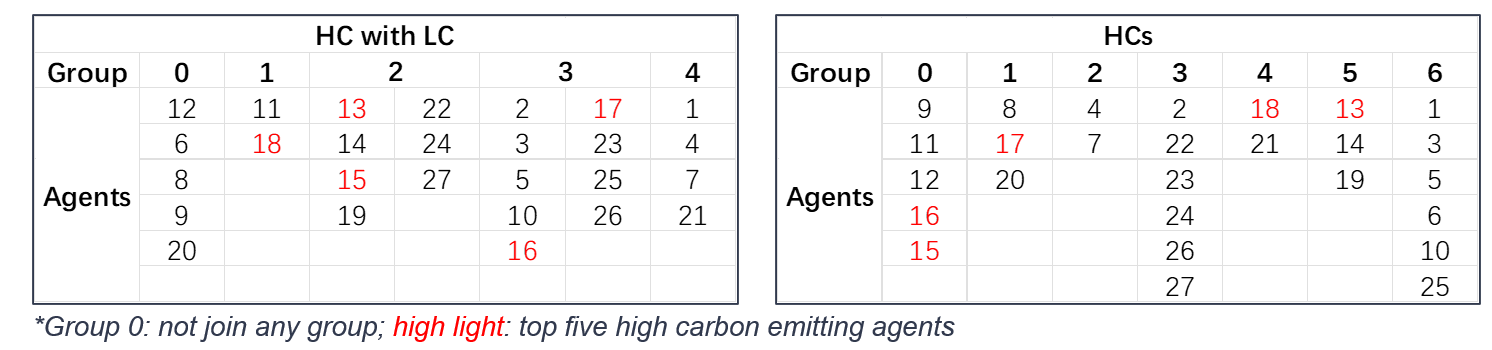}
  \caption{Group state during final time step.}
\end{figure}

\section{Conclusions}
\label{Conclusions}

In general, based on the Nordhaus' Climate Club [2] theory and real-world situation, this work proposed an approach to climate negotiations based on existing trade groups. Current research suggested that the model can inform real-world climate negotiations, but the underlying reasons remain to be discovered.

\section*{Supplementary Material}
For the implementation of the above settings, please refer to the relevant files in the Supplementary Materials.

\section*{Acknowledgments}
This work reused the AI4GCC code from MILA and Salesforce.

\section*{References}
\medskip

{
\small
[1] IPCC, 2023: Summary for Policymakers. In: Climate Change 2023: Synthesis Report.A Report of the Intergovernmental Panel on Climate Change. Contribution of Working Groups I, II and III to the Sixth Assessment Report of the Intergovernmental Panel on Climate Change [Core Writing Team, H. Lee and J. Romero (eds.)]. IPCC, Geneva, Switzerland.

[2] Nordhaus W. (2015) Climate Clubs: Overcoming Free-Riding in International Climate Policy. {\it American Economic Review} {\bf 105}(4):1339-70.

[3] Nordhaus W. (2018) Evolution of modeling of the economics of global warming: changes in the DICE model, 1992–2017. {\it Climatic Change} {\bf 148}(4):623–640.

[4] Tianyu Z, et al. AI for global climate cooperation: Modeling global climate negotiations, agreements, and long-term cooperation in rice-n, 2022. https://doi.org/10.48550/arXiv.2208.07004.

[5] Schippers M.L.,Wit W.De. (2022) Proposal for a Carbon Border Adjustment Mechanism, {\it Global Trade and Customs Journal} {\bf 17}(1):10-18

[6] He L-Y, Huang G. (2020) Tariff Reduction and Environment: Evidence from CAFTA and Chinese Manufacturing Firms. {\it Sustainability} {\bf 12}(5):2017. 

[7] Norris J. G7 pins hopes on ‘climate club’ as the saviour of 1.5C target, 2022. https://chinadialogue.net/en/climate/g7-pins-hopes-on-climate-club-as-the-saviour-of-1-5c-target.

[8] Eicke L,Weko S,Apergi M,Marian A. (2021) Pulling up the carbon ladder Decarbonization,dependence,and third-country risks from the European carbon border adjustment mechanism. {\it Energy Res Social Sci} {\bf 80}:102240
}

\end{document}